\documentclass[%
reprint,
superscriptaddress,
 amsmath,
 amssymb,
 aip,jcp,
]{revtex4-2}

\usepackage{mathrsfs}
\usepackage{mathtools}
\usepackage{bm}
\usepackage{subfigure}
\usepackage{graphicx}
\usepackage{dcolumn}
\usepackage{hyperref}

\usepackage{mathrsfs}
\usepackage{color}
\usepackage{braket}
\usepackage[version=4]{mhchem}
\usepackage[normalem]{ulem}

\begin{document}

\title{Spin/Phonon Dynamics in Single Molecular Magnets: II. spin/phonon entanglement}

\author{
Nosheen Younas$^{1,2,3}$,
Yu Zhang$^{2}$,
Andrei Piryatinski$^{2}$,
Eric R. Bittner$^{1,3}$
 }

\address{
$^{1}$
Department of Physics, University of Houston, Houston, Texas 77204, USA\\
$^{2}$Theoretical Division, Los Alamos National Laboratory, Los Alamos, New Mexico 87545, United States\\
$^{3}$Center for Nonlinear Studies, Los Alamos National Laboratory, Los Alamos, New Mexico 87545, United States
}

%
%


\begin{abstract}
We introduce a new quantum embedding method to explore spin-phonon interactions in molecular magnets. This technique consolidates various spin/phonon couplings into a limited number of collective degrees of freedom, allowing for a fully quantum mechanical treatment. By precisely factorizing the entire system into "system" and "bath" sub-ensembles, our approach simplifies a previously intractable problem, making it solvable on modest-scale computers. We demonstrate the effectiveness of this method by studying the spin relaxation and dephasing times of the single-molecule qubit \ce{VOPc(OH)8}, which features a lone unpaired electron on the central vanadium atom.  By using this mode projection method, we are able to perform numerical exact quantum dynamical calculation on this system which allows us to follow the flow of quantum information from the single spin qubit into the projected phonon degrees of freedom.  Our results demonstrate both the utility of the method and suggest how one can engineer the environment as to further optimize the quantum properties of a qubit system.
\end{abstract}


\maketitle


\section{Introduction}
Single molecular magnets (SMMs) and metal-organic frameworks (MOFs) have garnered considerable interest because of their potential in quantum information processing and their ability to maintain extended coherence times. However, these systems experience spin dephasing because of interactions and couplings with the molecular framework's vibrational motions.

Amongst the various MOF architectures, vanadyl phthalocyanine (VOPc) is a prominent candidate for room-temperature quantum computing
using the unpaired $d_{xy}$ as a prototypical 
qubit owing to its long $T_1$ and $T_2$ relaxation times, 
which are largely determined by weak interaction 
between the spin and vibrational degrees of freedom.
~\cite{Atzori:2016aa, Bonizzoni:2017aa, Malavolti:2018aa, Aziz:2011aa}  While
first-principles calculations, particularly density functional theory (DFT), have become essential in studying the electronic and magnetic properties of molecular magnets~\cite{ja061798a, Timco:2009tc}
challenges remain in terms of accurately modeling the
dynamical processes due to the high computational cost and complexity of spin-phonon interactions, which involve a large number of degrees of freedom.  Even though the
spin/phonon coupling is weak and can 
be treated as a perturbation, it is interesting to ascertain the mechanism by which energy and quantum 
information flows irreversibly from the central spin into the environment. 

In our accompanying paper, Ref.\citenum{paper1}, we introduced a systematic projection/embedding scheme to analyze spin-phonon dynamics in molecular magnets. By consolidating all spin/phonon couplings into a few collective degrees of freedom that can be treated fully quantum mechanically, this approach transforms a numerically intractable problem into one manageable one on modest-scale computers using an exact factorization of the complete system into ``system'' and ''bath'' sub-ensembles in which very few collective modes of the bath couple directly to the quantum system. 
The projected modes themselves carry
physical importance, since all of the
spin/phonon coupling is concentrated
in these collective degrees of freedom. 
Using parameters derived from \textit{ab initio} methods, we applied this approach to calculate the electronic spin relaxation and dephasing times for the single-molecule qubit \ce{VOPc(OH)8}, which contains a single unpaired electron on the central vanadium atom. This development provides a useful tool for simulating spin relaxation in complex environments with significantly reduced computational complexity.\cite{Yang:2015aa,Pereverzev:2009aa,Thouin:2019aa,Yang:2017aa}

Although the focus of Ref.~\citenum{paper1} was in the identification of the optical modes and the computation of
the $T_1$ and $T_2$ relaxation and dephasing times
for \ce{VoPC}, here we extend the ``system'' to include the bath modes identified by the projection method
within a unified spin + phonon dynamical scheme in order
to map the flow of quantum information 
from the spin qubit into the environment.  
We find for this system that the spin rapidly becomes 
entangled with at least one of the collective 
modes identified by our approach and that this mode
provides a conduit for energy and quantum information 
loss from the qubit.  We speculate that if spin/phonon coupling can be manipulated by chemical modification
of the surrounding molecular scaffold,
it may be possible to enhance the $T_1$ and $T_2$ times
of other SMM systems.

\section{Theory}
\subsection{System/bath decomposition}
We begin with a brief summary of the approach developed
in Ref.~\citenum{paper1}.  Starting from the Zeeman 
Hamiltonian describing the spin-orbit coupling for an electron in a molecular orbital
\begin{align}
    H = \beta  \hat{\mathbf{S}} \cdot \mathbf{\stackrel{\leftrightarrow}{g}}  \cdot \mathbf{B}
    +\frac{1}{2}\sum_n^{3N-6} p_n^2 + \omega_n^2x_n^2
    \label{eq:1}
\end{align}
The first term is the usual Zeeman 
interaction between an external magnetic
field $\bf B$ and electronic spin $\bf\hat  S$
and $\beta = \mu_B/\hbar$.
The two are coupled via the 
g-tensor, $\stackrel{\leftrightarrow}{g}$,
defined at some 
optimized molecular geometry of the
molecular system. 
Likewise, the $3N-6$ normal modes $\{x_n\}$ and 
associated frequencies $\{\omega_n\}$ are determined 
at the optimized molecular geometry. 
The coupling between the spin and nuclear motions
enters by expanding the g-tensor in terms of 
nuclear displacements along the normal modes. 
Although nuclear displacements
do not appear directly in this expression, $\bm{g}$  depends upon the local molecular geometry
via second-order perturbation theory
\begin{align}
    g_{\alpha\beta} = g_o + \sum_e \frac{\langle \psi_g|\hat L_\alpha|\psi_e\rangle
    \langle \psi_e| \hat L_\beta|\rangle}
    {E_e-E_g},
\end{align}
where $g_o$ is the g-value for a 
free electron and $\hat L_\alpha$ 
are the orbital angular momentum operators coupling the ground electronic state $|\psi_g\rangle$ to 
the unoccupied electronic states $|\psi_e\rangle$ taken at the 
equilibrium nuclear geometry.  
Of the terms entering this expression, the energy gap is most sensitive to fluctuations in
equilibrium geometry.  Expanding about the 
equilibrium geometry along the normal coordinates gives
 \begin{align}
 H = 
 \underbrace{
\left(\beta \hat{\mathbf{S}} \cdot \mathbf{\stackrel{\leftrightarrow}{g}}  \cdot \mathbf{B}\right)_o}_{\rm Zeeman\, term} + \underbrace{\sum_{\alpha,n} g'_{\alpha n} \hat \sigma_\alpha x_n}_{\rm Spin/phonon\, interaction}
+
\underbrace{\frac{1}{2}\sum_n^{3N-6} p_n^2 + \omega_n^2x_n^2}_{\rm normal\, modes}
\label{eq:3}
\end{align}   
where $g'_{\alpha n}$ denotes the derivative coupling
between the $\alpha = x,y,z$ spin operator and the
$n$-th normal model.

\begin{widetext}
In our previous work\cite{paper1}, we further
partition the normal modes into ``primary'' and
residual modes by finding collective degrees of freedom that optimize the spin / phonon coupling by performing a single value decomposition in $g'$ \textit{ viz.}
\begin{equation}
    \bm{g}' = [g'_{\alpha n}]_{n_s\times n_q} = \bm U_{n_s \times n_s} \cdot \bm \Sigma_{n_s\times n_q} \cdot \bm V^\dagger_{n_q \times n_q}
\end{equation}
where $\bm \Sigma$ is a matrix of singular values with at most $\min(n_s, n_q)$ nonzero singular values. 
The ${\bm U}$ and $\bm V$ 
allow us define a series of 
canonical transformations 
which brings the full spin+phonon 
Hamiltonian into the following form
 \begin{align}
H \approx& 
\underbrace{
\left(\beta \hat{\mathbf{S}} \cdot \mathbf{\stackrel{\leftrightarrow}{g}}  \cdot \mathbf{B}\right)_o}_{\rm Zeeman\, term} + \underbrace{\sum_{\alpha,k\in{\cal P}} g'_{\alpha k} \hat \sigma_\alpha X_k}_{\rm Spin/phonon\, interaction}
\nonumber
\\
&+
\overbrace{
\underbrace{
\sum_{k\in {\cal P}} \frac{1}{2}\left(P_k^2+ \omega_k^2 X_k^2 \right)}_{\rm primary\, modes}
+
\underbrace{\sum_{q\in {\cal Q}} \frac{1}{2}\left(
{P_q^2} + \omega_q^2 Y_q^2 \right)
}_{\rm residual\, modes}
+
\underbrace{\sum_{k\in {\cal P},q\in {\cal Q}} \gamma_{kq} \hat{X}_k \hat{Y}_q}_{linear\, couplings}
}^{\rm normal\, modes} \label{eq:5}
\end{align}   

The last three terms in this expression
define the Hamiltonian for all the
vibrational degrees of freedom, written
in terms of the projected ``primary'' and ``residual'' modes in which the
``primary'' modes are coupled directly to the
spin and the ``residual'' modes are only
coupled to the ``primary'' modes, 
both of which we write in terms of quantum
position and momentum operators. 

Formally, Eq.\ref{eq:3} and Eq.~\ref{eq:5} are
identical. However, a full quantum treatment of the
system remains intractable.   Consequently, 
we will restrict the quantum evolution to the 
spin + primary $\{X_k\}$ modes and tracing over the residual degrees of freedom and defining the $1_k \to 0_k$ relaxation 
rate for the $k$-th mode as\cite{Kenkre:1994aa}
\begin{align}
    \frac{1}{T_{vib,k}} &= \frac{1}{\hbar^2}
    |\langle 0_k|X_k|1_k\rangle|^2
    \sum_q
    \int_{-\infty}^{+\infty}d\tau     e^{i\omega_k \tau}\gamma_{kq}^2\langle
    Y_q(\tau)Y_q(0)\rangle\\
    &=
    \frac{1}{2\hbar\omega_k}
    \sum_q
    \int_{-\infty}^{+\infty}    e^{i\omega_k \tau}
    \gamma_{kq}^2\langle
    Y_q(\tau)Y_q(0)\rangle d\tau
    \label{eq:decay_rates}
\end{align}where
\begin{align}
    \langle Y_q(t)Y_q(0)\rangle &= \frac{1}{Z_Q}
    {\rm Tr}_Q\left[e^{-\beta H_Q}e^{+iH_Qt/\hbar}Y_q
    e^{-iH_Qt/\hbar}Y_q
    \right]\nonumber \\
    &=\frac{\hbar}{2\omega_q}((1+n(\omega_q))e^{-i\omega_q t}
    +n(\omega_q)e^{+i\omega_q t}),
\end{align}
where $n(\omega)$ is the Bose-Einstein population for an oscillator with frequency $\omega$. 
Integrating over time 
\begin{align}
    \frac{1}{T_{vib,k}}= 
    \frac{\pi}{\hbar\omega_k}
    \sum_q \frac{\hbar \gamma_{kq}^2}{2\omega_q} ((1+n(\omega_q))\delta(\omega_k-\omega_q) + n(\omega_q)\delta(\omega_q-\omega_k))
\end{align}
The relaxation rates ($k_i$) for all three primary modes are presented in Fig.~\ref{fig:inverse_t1} for a temperature range of 10 to 300 K. The results indicate that all three modes have nearly identical decay rates at low temperatures. With increasing temperature, all decay rates increase. At intermediate temperature (around 100 K), modes 2 and 3 start to exhibit a higher gain in the decay rate, and their decay rate surpasses that of mode one. This trend continues until room temperature, with mode two showing the highest decay rate (hence, the shortest $T_1$). This is due to populating bath modes that interact with modes 2 and 3 relatively more strongly than mode 1.
\begin{figure}
    \centering
    \includegraphics[width=0.55\textwidth]{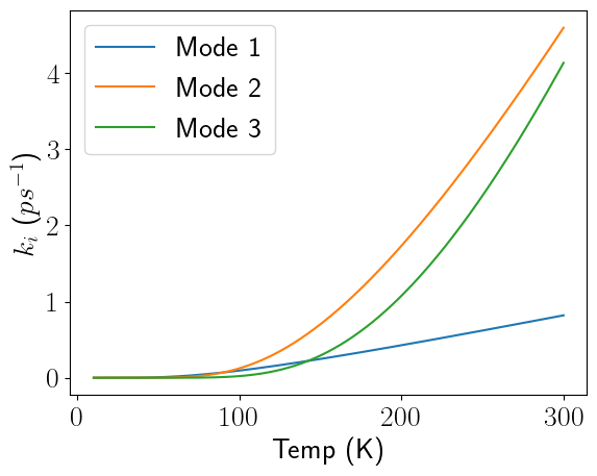}
    \caption{Vibrational relaxation rates, $k_i$, for all three primary modes, as given by Eq.~\ref{eq:decay_rates}. All three decay rates are nearly identical at low temperatures. At intermediate temperature, modes 2 and 3 show increased gain in decay rate that continues till 300K. Consequently, at 300 K, mode two shows the highest decay rate, followed by modes 3 and 1.}
    \label{fig:inverse_t1}
\end{figure}

\subsection{Dynamics}
We now consider the 
full quantum dynamics of a single spin
coupled directly to
the projected modes, which in turn are 
coupled to a completely dissipative 
environment.  For this we, employ
the Lindblad approach\cite{Lindblad:1976aa}
\begin{align}
    \dot\rho(t)&=-\frac{i}{\hbar}[H_s,\rho(t)] 
    +
    \sum_{k,\alpha=\pm} \frac{1}{2} \left[2 C_{k\alpha} \rho(t) C_{k\alpha}^\dagger - \{\rho(t),C_{k\alpha}^\dagger C_{k\alpha}\}\right]
\end{align}
in which $H_s$ contains the Zeeman term
for the spin and its coupling to the three 
primary modes
\begin{align}
    H_s =\beta\sum_{\alpha=x,y,z} B_\alpha g_\alpha
    \hat \sigma_\alpha
    + 
    \sum_{\alpha k}g'_{\alpha k}\sigma_\alpha(a_k^\dagger + a_k)
    +\sum_{k}\hbar\omega_k (a^\dagger_k a_k + 1/2).
\end{align}
Coupling to the residual modes is captured by the
Lindblad relaxation operators as given by
\begin{align}
       C_{k,-} =\sqrt{\frac{n(\omega_k)+1}{T_{vib,k}} }a \,\,\,\& \,\,\,
       C_{k,+} =\sqrt{\frac{n(\omega_k)}{T_{vib,k}} }a^\dagger
\end{align}
corresponding to vibrational de-excitation 
and excitation of the $k$-th primary modes 
due to the thermal fluctuations of the 
residual modes.

\subsection{Isolated system dynamics}
We first consider the dynamics
in the absence of the residual bath.
The transition frequency between the two 
possible spin states is set by the 
external magnetic field, typically between 1-10T
Under these conditions, the transition frequency between spin levels is much smaller compared to the frequencies of the system phonons. Consequently, the spin evolves on a much slower 
timescale than the phonons. This is 
advantageous for molecular qubits since
this detuning leads much longer coherence 
and relaxation times.
Additionally, the couplings ($g_{ik}$) are directly proportional to the applied magnetic field; therefore, a lower field (e.g. 1 T) will result in a significantly smaller spin-phonon interaction and quantum exchange.
For purposes of our 
model, we can increase the applied magnetic field to 200 Tesla, thus bringing the spin transition frequency to 185.51$cm^{-1}$. This is close to the first system phonon frequency 
and allows us to examine 
the quantum dynamics of this system
and the role of spin/phonon entanglement.

\subsection{Spin with Isolated Projected Modes}

We construct a system with spin and three projected modes without the addition of any interaction with the bath.
Initially, the spin is excited, while all projected phonon modes start in the ground state. In number state representation,
\begin{equation}
|m;\{\ n_1, \ n_2, \ n_3\}\rangle = |+\frac{1}{2};\{0_1, 0_2, 0_3\}\rangle.
\end{equation}
This setup allows for a controlled exploration of the dynamics and interactions of the isolated system.
Since the spin-phonon coupling is weak, the populations of these states should remain close to their initial values. Fig.~\ref{fig:subsystem2} shows the results for this simulation with populations given in terms of the deviation from their initial value, {\em i.e.} $\delta \rho = \rho(t)-\rho(0)$. The population of the excited state for spin is shown in panel (a), while that of the ground state of mode 1 is shown in panel (b). Both show nearly identical dynamics, a consequence of the similar energies (or frequencies) of these 2-level systems. The dynamics of modes 2 and 3, shown in panels (c) and(d),) is markedly different.
Furthermore, mode 1 shows beats (similar to spin), mode 2 exhibits simple oscillations, and mode 3 undergoes high-frequency oscillations on top of a low-frequency wave. Relatively, the amplitude of oscillations for mode 1 is larger than for mode 2, with mode 3 displaying the smallest amplitude.

\begin{figure*}[htb]
     \centering
     \includegraphics[width=0.7\textwidth]{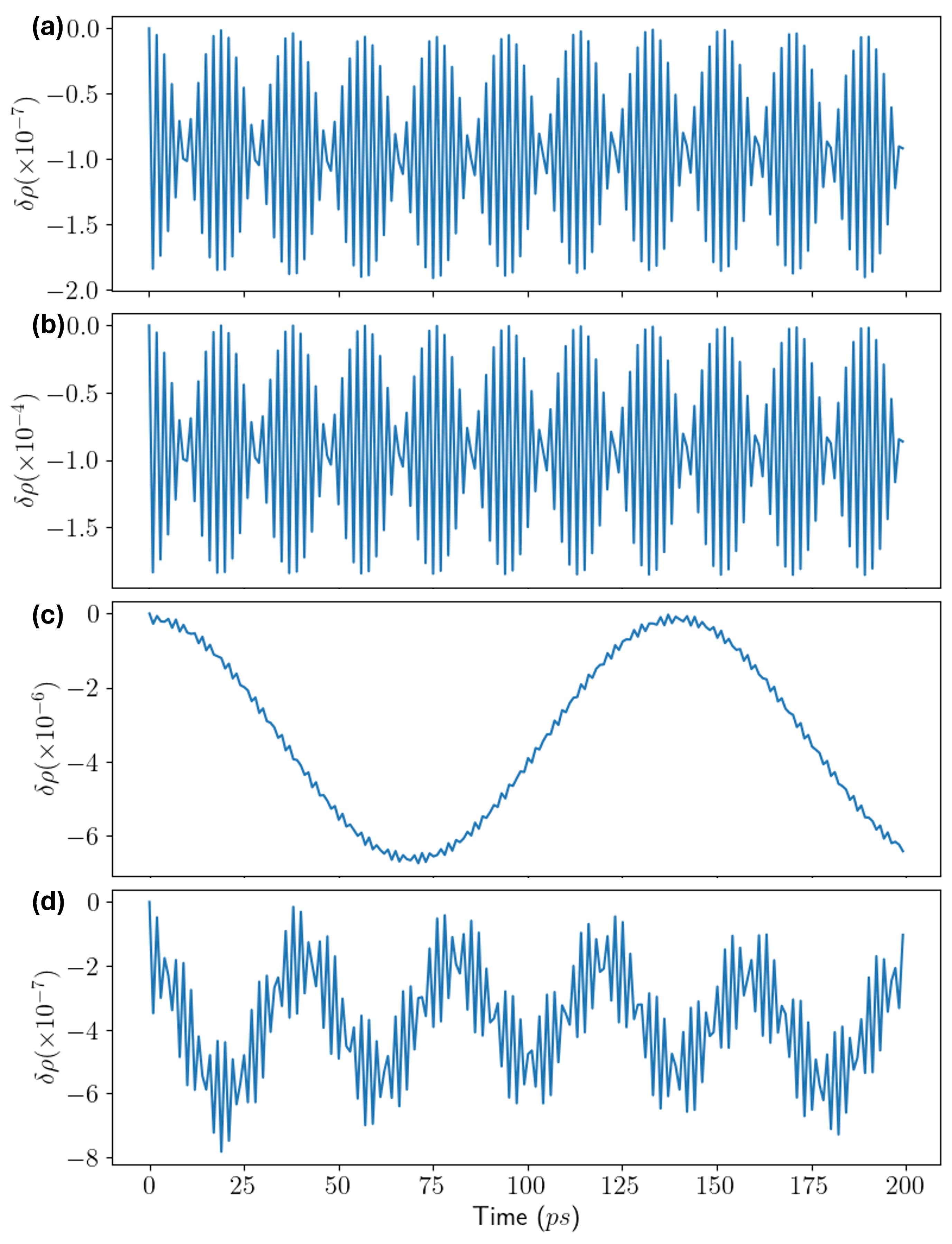}
        \caption{Population dynamics of spin and three projected modes represented using $\delta \rho = \rho(t)-\rho(0)$. The initial state is given by $|spin, phonon_1, phonon_2, phonon_3\rangle = |1, 0, 0, 0\rangle$. Panel (a) shows the excited state population of the spin. Panels (b), (c), and (d) show ground state populations for modes 1, 2, and 3, respectively with $\delta \rho = \rho_{00}-1$. Since the interaction between spin and phonons is weak, these populations are very close to their initial value throughout the evolution.
        }
        \label{fig:subsystem2}
\end{figure*}

\end{widetext}
\subsection{Thermalized Projected Modes }

\subsubsection{Including Spin}\label{results}
Finally, we simulate the evolution of the spin with all three projected modes, including the effect of the thermal bath calculated previously. Initially, the spin is excited, while the phonons are in their ground state. 
The combined system evolved at 65 K, a typical temperature for liquid-nitrogen-cooled experiments. 
The lifetimes of system modes at this temperature are $43.3$, $127.1$, and $2879.1$ ps for mode 1, 2, and 3, respectively. 
The results are presented in Fig. reffig: simulation4b, which shows the $ rho_11$ element of spin and $ rho_00$ elements of modes 1 to 3 (marked by $ p1, p2$, and $ p3$). Since modes 1 and 2 exhibit relatively short lifetimes compared to mode 3, their populations reach thermodynamic equilibrium much faster than mode 3's. On the other hand, the spin and mode 3 are still decaying by the end of the simulation.

\begin{figure}
    \centering
    \includegraphics[width=0.55\textwidth]{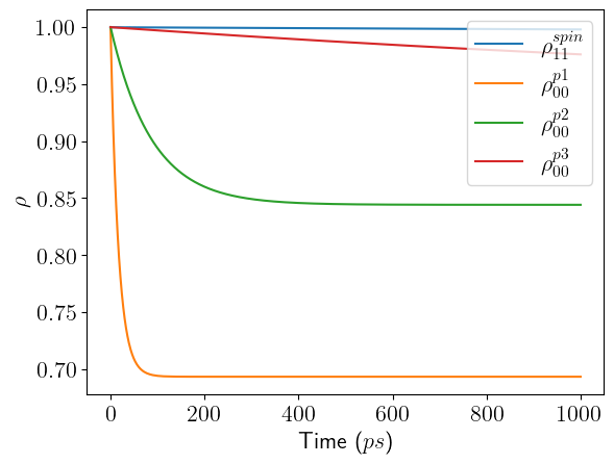}
    \caption{Population dynamics for the spin and three system-phonons at 65 K. The total system starts in $|1,0,0,0\rangle$ state. Consequently, excited state population is shown for spin ($\rho_{11}^{spin}$) and ground state populations are shown for phonons ($\rho_{00}^{pi}$). All populations are represented by elements of density matrices for corresponding subsystems. Notice that the populations of modes 1 and 2 have reached their thermodynamic equilibrium but those of mode 3 and spin are still relaxing by the end of simulation. }
    \label{fig:simulation4b}
\end{figure}
\begin{figure}
    \centering
     \includegraphics[width=0.5\textwidth]{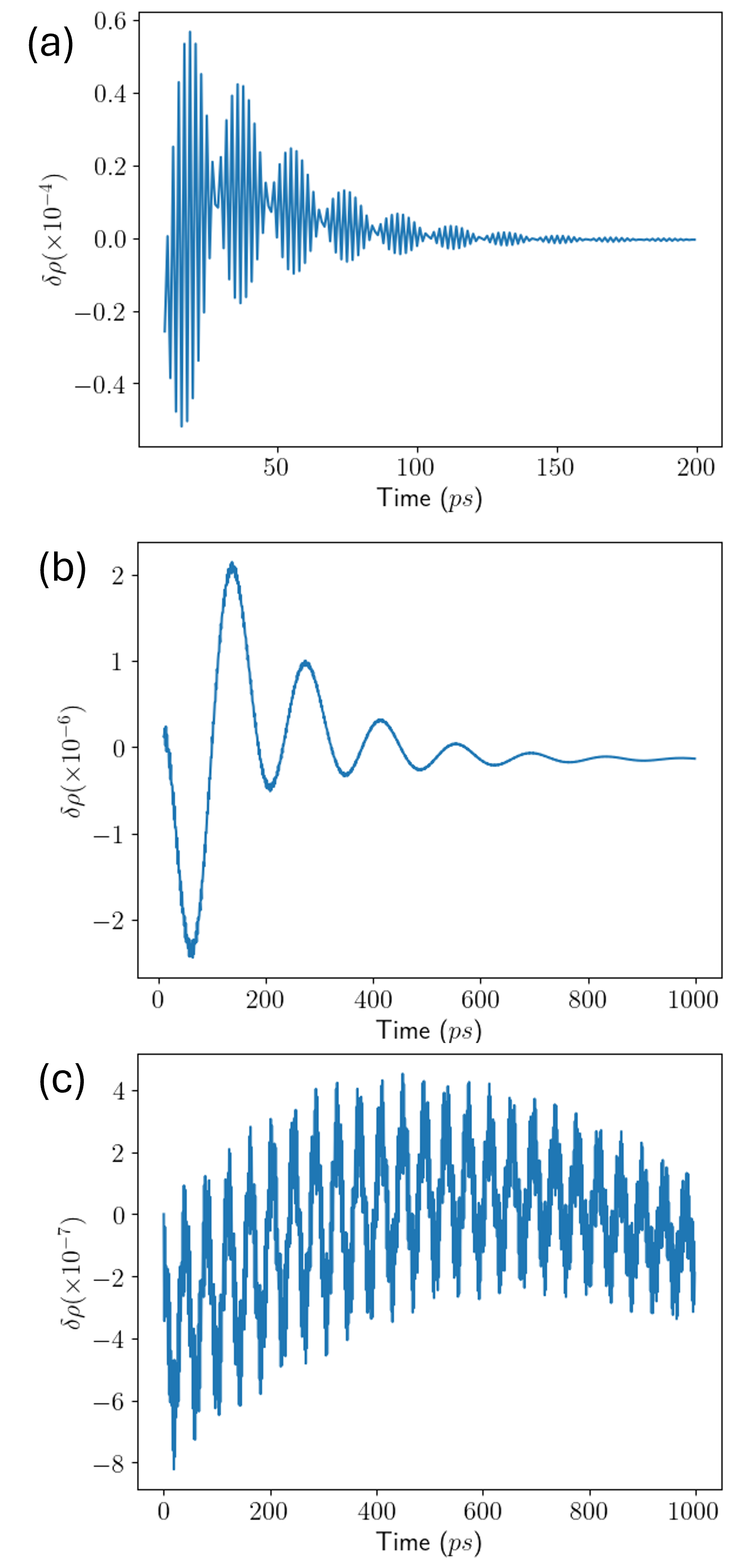}
    \caption{Population dynamics of the ground state for (a) mode 1, (b) mode 2, and (c) mode 3, at 65 K starting from $|1,0,0,0\rangle$. The exponential thermal decay for each mode has been removed individually for clarity such that $\delta \rho = \rho(t)-\rho^{th}(t)$. We notice that these dynamics are identical to those observed in Test Case 1, despite an overall thermal decay.}
    \label{fig:simulation4mode1}
\end{figure}

At first glance, all three modes seem to be decaying thermodynamically, following $\rho^{th}(t)$. However, intricate dynamics can be resolved by removing this exponential decay profile induced by thermalization.
We remove the exponential decay profile for mode 1 by introducing, 
$$
\delta \rho = \rho(t) - \rho^{th}(t),
$$ 
The results are presented in Fig.~\ref{fig:simulation4mode1}(a).
This enhanced perspective highlights that the first system phonon undergoes coherent oscillations primarily driven by its interaction with the spin. These oscillations also indicate beats with tens of picoseconds time period.

A similar evolution for mode 2 is presented in Fig.~\ref{fig:simulation4mode1}(b) indicating population oscillations with a time period of about 200 ps. Likewise, Fig.~\ref{fig:simulation4mode1}(c) shows the ground state evolution for mode 3, displaying fast oscillations at a
period of $\approx 20$ps. The dynamics of all three modes are qualitatively similar to the case of an isolated system (Fig.~\ref{fig:subsystem2}), although thermalization is due to interaction with the bath.

\subsection{Spin-phonon Correlation}

To quantify the degree of entanglement between spin and three projected modes, we analyze mutual information between the spin and the phonon degrees of freedom as the total system evolves. 
Mutual information measures correlation between parts of a composite quantum mechanical system. 
Consider a bipartite quantum system ($A\otimes B$), mutual information $I(A:B)$ reflects the extent of shared information between the two subsystems $A$ and $B$. It is mathematically defined as 
\begin{equation}
\label{eq:mutual_info}
    I(A:B) = S(A) + S(B) - S(A\otimes B)
\end{equation}
where, $S(A)$, $S(B)$, and $S(A\otimes B)$ represent von Neumann entropies of the sub-systems $A$, $B$, and the composite system $A\otimes B$, respectively. $I(A:B)$ serves as a measure of the correlations between components $A$ and $B$, capturing both classical and quantum aspects. This measure is useful in quantum information theory, as it indicates the level of entanglement and the degree to which the state of one subsystem informs about the state of the other. High mutual information signifies strong correlations, which can be harnessed in quantum computing and quantum communication protocols to improve performance and security. In the present context, we use $I(A:B)$ to measure the degree of entanglement between the single spin qubit and each of the three possible projected modes. 

\begin{figure}
    \centering
    \includegraphics[width=0.5\textwidth]{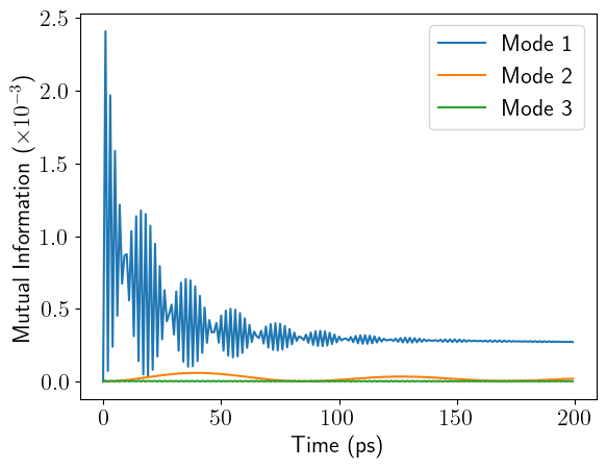}
    \caption{Mutual information between the central
    spin and each of the three phonon degrees of freedom. Mode 1  displays fast oscillations with slow beats, which are absent in modes 2 and 3.}
    \label{fig:mutual_info}
\end{figure}

To investigate the correlation between the spin and a specific system mode, we first trace over the remaining two modes and then apply the mutual information formulation (Eq.~\ref{eq:mutual_info}) using von Neumann entropy. The results are shown in Fig.~\ref{fig:mutual_info}.
Initially, the spin and phonon degrees of freedom are not entangled.
However, we can see that the first mode shows the most correlation with spin, followed by the second and third modes. This means that the quantum 
information ``flows'' from the spin state through mode 1 more readily than through the other two modes.
This relatively higher degree of correlation between the spin and the first mode is reasonable since both are nearly resonant, even though the first mode does not show the highest coupling with spin via any Pauli operators.

\section{Summary}
The reduction of phonon degrees of freedom via SVD-mode projection has enabled the full quantum treatment of the spin-phonon system. We investigate the spin-phonon interaction by building models with and without a thermal bath.  These population oscillations preserve their character even in a thermal bath, a signature of their quantum nature.

\label{sec:summary}

This work introduced a novel quantum embedding approach to study spin-phonon dynamics in molecular magnets. Our method efficiently explores the spin-lattice relaxation process with high accuracy. We applied this approach to investigate the spin dynamics in single molecular magnets. Leveraging this method, we have deepened our understanding of the physics underlying spin relaxation in molecular magnets, which is essential for enhancing their performance across various applications. Furthermore, this study lays the groundwork for rationalising new materials with customized magnetic properties suitable for quantum information processing and other applications.

As a future direction, we plan to extend the quantum embedding method to encompass spin-spin interactions in molecular magnets. This expansion will provide insights into the interplay between spin-spin and spin-phonon interactions and elucidate the roles these interactions play in shaping the overall relaxation dynamics of molecular qubits. Exploring spin-spin interactions will deepen our understanding of these intricate systems and aid in creating superior materials for quantum computing and other applications involving magnetism. Such an extension promises to significantly widen the quantum embedding approach's scope and utility, enhancing its value as a research tool in the molecular magnet domain.

\section*{Acknowlegments}
The research presented in this article was supported by the LANL LDRD program (number 20220047DR). 
LANL is operated by Triad National Security, LLC, for the National Nuclear Security Administration of the U.S. Department of Energy (contract no. 89233218CNA000001). We thank the LANL Institutional Computing (IC) program for access to HPC resources. 
The work at the University of Houston was funded in part by the National Science Foundation (CHE-2404788)  
and the Robert A. Welch Foundation (E-1337).
This work was performed, in part, at the Center for Integrated Nanotechnologies, an Office of Science User Facility operated for the U.S. Department of Energy (DOE) Office of Science by Los Alamos National Laboratory (Contract 89233218CNA000001) and Sandia National Laboratories (Contract DE-NA-0003525).

\section*{Data Availability Statement}
The data supporting this study's findings are available from the corresponding author upon reasonable request.

\section*{Author Contribution Statement}
The authors confirm their contribution to the paper as follows: study conception and design: ERB, YZ, and AP; data collection and simulations: NY; analysis and interpretation of results: NY, YZ, AP, ERB; draft manuscript preparation: NY, ERB. All authors reviewed the results and approved the final version of the manuscript.


\begin{thebibliography}{99}

\bibitem{Atzori:2016aa}
Atzori M, Tesi L, Morra E, Chiesa M, Sorace L, Sessoli R. 2016
  Room-Temperature Quantum Coherence and Rabi Oscillations in Vanadyl
  Phthalocyanine: Toward Multifunctional Molecular Spin Qubits. {\em J Am Chem
  Soc} \textbf{138}, 2154--2157.
(\href{http://dx.doi.org/10.1021/jacs.5b13408}{10.1021/jacs.5b13408})

\bibitem{Bonizzoni:2017aa}
Bonizzoni C, Ghirri A, Atzori M, Sorace L, Sessoli R, Affronte M. 2017
  Coherent coupling between Vanadyl Phthalocyanine spin ensemble and microwave
  photons: towards integration of molecular spin qubits into quantum circuits.
  {\em Scientific Reports} \textbf{7}, 13096.
(\href{http://dx.doi.org/10.1038/s41598-017-13271-w}{10.1038/s41598-017-13271-w})

\bibitem{Malavolti:2018aa}
Malavolti L, Briganti M, H{\"a}nze M, Serrano G, Cimatti I, McMurtrie G, Otero
  E, Ohresser P, Totti F, Mannini M, Sessoli R, Loth S. 2018  Tunable
  Spin-Superconductor Coupling of Spin 1/2 Vanadyl Phthalocyanine Molecules.
  {\em Nano Lett.} \textbf{18}, 7955--7961.
(\href{http://dx.doi.org/10.1021/acs.nanolett.8b03921}{10.1021/acs.nanolett.8b03921})

\bibitem{Aziz:2011aa}
Nadeem S, Khan MN, Muhammad N, Ahmad S. 2019  {Mathematical analysis of
  bio-convective micropolar nanofluid}. {\em Journal of Computational Design
  and Engineering} \textbf{6}, 233--242.
(\href{http://dx.doi.org/10.1016/j.jcde.2019.04.001}{10.1016/j.jcde.2019.04.001})

\bibitem{ja061798a}
Neese F. 2006  Importance of direct spin spin coupling and spin-flip
  excitations for the zero-field splittings of transition metal complexes: a
  case study. {\em J Am Chem Soc} \textbf{128}, 10213--10222.
(\href{http://dx.doi.org/10.1021/ja061798a}{10.1021/ja061798a})

\bibitem{Timco:2009tc}
Timco GA, Carretta S, Troiani F, Tuna F, Pritchard RJ, Muryn CA, McInnes EJL,
  Ghirri A, Candini A, Santini P, Amoretti G, Affronte M, Winpenny REP. 2009
  Engineering the coupling between molecular spin qubits by coordination
  chemistry. {\em Nature Nanotechnology} \textbf{4}, 173--178.
(\href{http://dx.doi.org/10.1038/nnano.2008.404}{10.1038/nnano.2008.404})

\bibitem{paper1}
Younas N, Zhang Y, Piryatinski A, Bittner ER. submitted  Spin/Phonon Dynamics
  in Single Molecular Magnets: I. quantum embedding. {\em Proc. Roy. Soc. A}.

\bibitem{Yang:2015aa}
Yang X, Bittner ER. 2015  {Computing intramolecular charge and energy transfer
  rates using optimal modes}. {\em The Journal of Chemical Physics}
  \textbf{142}, 244114.
(\href{http://dx.doi.org/10.1063/1.4923191}{10.1063/1.4923191})

\bibitem{Pereverzev:2009aa}
Pereverzev A, Bittner ER, Burghardt I. 2009  {Energy and charge-transfer
  dynamics using projected modes}. {\em The Journal of Chemical Physics}
  \textbf{131}, 034104.
(\href{http://dx.doi.org/10.1063/1.3174447}{10.1063/1.3174447})

\bibitem{Thouin:2019aa}
Thouin F, Srimath~Kandada AR, Valverde-Ch{\'a}vez DA, Cortecchia D, Bargigia I,
  Petrozza A, Yang X, Bittner ER, Silva C. 2019  Electron--Phonon Couplings
  Inherent in Polarons Drive Exciton Dynamics in Two-Dimensional Metal-Halide
  Perovskites. {\em Chemistry of Materials} \textbf{31}, 7085--7091.
(\href{http://dx.doi.org/10.1021/acs.chemmater.9b02267}{10.1021/acs.chemmater.9b02267})

\bibitem{Yang:2017aa}
Yang X, Keane T, Delor M, Meijer AJHM, Weinstein J, Bittner ER. 2017
  Identifying electron transfer coordinates in donor-bridge-acceptor systems
  using mode projection analysis. {\em Nature Communications} \textbf{8},
  14554.
(\href{http://dx.doi.org/10.1038/ncomms14554}{10.1038/ncomms14554})

\bibitem{Kenkre:1994aa}
Kenkre VM, Tokmakoff A, Fayer MD. 1994  {Theory of vibrational relaxation of
  polyatomic molecules in liquids}. {\em The Journal of Chemical Physics}
  \textbf{101}, 10618--10629.
(\href{http://dx.doi.org/10.1063/1.467876}{10.1063/1.467876})

\bibitem{Lindblad:1976aa}
Lindblad G. 1976  On the generators of quantum dynamical semigroups. {\em
  Communications in Mathematical Physics} \textbf{48}, 119--130.
(\href{http://dx.doi.org/10.1007/BF01608499}{10.1007/BF01608499})

\end{thebibliography}

\vskip2pc

\end{document}